# First astrophysical detection of the helium hydride ion (HeH$^+$)

Rolf Güsten[1], Helmut Wiesemeyer[1], David Neufeld[2], Karl M. Menten[1], Urs U. Graf[3], Karl Jacobs[3], Bernd Klein[1,4], Oliver Ricken[1], Christophe Risacher[1,5], Jürgen Stutzki[3]

[1] Max-Planck Institut für Radioastronomie, Auf dem Hügel 69, 53121 Bonn, Germany
[2] The Johns Hopkins University, 3400 North Charles St., Baltimore MD 21218, USA
[3] I. Physikalisches Institut, Universität zu Köln, Zülpicher Str. 77, 50937 Köln, Germany
[4] University of Applied Sciences Bonn-Rhein-Sieg, 53757 Sankt Augustin, Germany
[5] Institut de Radioastronomie Millimétrique, 300 Rue de la Piscine, 38400 Saint-Martin-d'Hères, France

During the *dawn of chemistry*[1,2] when the temperature of the young Universe had fallen below ~4000 K, the ions of the light elements produced in Big Bang nucleosynthesis recombined in reverse order of their ionization potential. With its higher ionization potentials, He$^{++}$ (54.5 eV) and He$^+$ (24.6 eV) combined first with free electrons to form the first neutral atom, prior to the recombination of hydrogen (13.6 eV). At that time, in this metal-free and low-density environment, neutral helium atoms formed the Universe's first molecular bond in the helium hydride ion HeH$^+$, by radiative association with protons (He + H$^+$ → HeH$^+$ + hν). As recombination progressed, the destruction of HeH$^+$ (HeH$^+$ + H → He + H$_2^+$) created a first path to the formation of molecular hydrogen, marking the beginning of the *Molecular Age*.

Despite its unquestioned importance for the evolution of the early Universe, the HeH$^+$ molecule has so far escaped unequivocal detection in interstellar space. In the laboratory the ion was discovered as long ago as 1925[3], but only in the late seventies was the possibility that HeH$^+$ might exist in local astrophysical plasmas discussed[4,5,6,7]. In particular, the conditions in planetary nebulae were shown to be suitable for the production of potentially detectable HeH$^+$ column densities: the hard radiation field from the central hot white dwarf creates overlapping Strömgren spheres, where HeH$^+$ is predicted to form, primarily by radiative association of He$^+$ and H.

With the GREAT spectrometer[8,9] on board SOFIA[10] the HeH$^+$ rotational ground-state transition at λ149.1 μm is now accessible. We report here its detection towards the planetary nebula NGC7027. The mere fact of its proven existence in nearby interstellar space constrains our understanding of the chemical networks controlling the formation of this very special molecular ion.

The planetary nebula NGC7027 seems a natural candidate for a search for HeH$^+$: The nebula is very young (with a kinematic age of only 600 years)[11], and its shell of released stellar material is still rather compact and dense. The central star is one of the hottest known ($T_{eff}$ ~190,000 K) and is very luminous (1.0 ×10$^4$ L$_\odot$)[12]. Under these conditions the Strömgren spheres are not fully developed yet, and the hard intense radiation field drives ionization fronts into the molecular envelope. The He$^+$ Strömgren sphere will extend slightly beyond the H$^+$ zone, and it is in this thin overlap layer where HeH$^+$ will be produced. Detailed calculations[13] led to predictions for the intensities of the v =1 – 0 R(0) and P(2) ro-vibrational transitions in the near-infrared that were not confirmed despite of deep searches[14,15], however. Observations[16] with the ISO Long Wavelength Spectrometer of the pure rotational $J = 1 – 0$ ground-state transition at 149.137 µm were impaired by the limited resolving power of the spectrometer (Δλ = 0.6 µm) that did not allow the HeH$^+$ transition to be separated from the nearby Λ-doublet of CH at 149.09 and 149.39 µm.

Very debatable tentative detections of HeH$^+$ have been reported towards SN1987A[17] and a high-redshift quasar[18], all non-confirmed and suggested to be considered as upper limits. This lack of direct evidence of the very existence of the molecule has called into question our understanding of the underlying reaction networks[19,20] for local plasmas, that might ultimately invalidate current models of the early universe.

The operation of the GREAT[9] heterodyne spectrometer on board the Stratospheric Observatory for Infrared Astronomy[10] has now opened new opportunities. While the HeH$^+$ $J =1 – 0$ transition at 149.137 µm (2010.183873 GHz[21]) cannot be observed from ground-based observatories, skies become transparent during high-altitude flights with SOFIA. The latest advances in terahertz technologies have enabled the operation of the high-resolution spectrometer upGREAT[22] at frequencies above 2 THz, allowing the HeH$^+$ $J =1 – 0$ line to be targeted. This heterodyne instrument's resolving power, λ/Δλ ~10$^7$, permits the HeH$^+$ $J =1 – 0$ line to be distinguished unambiguously from other, nearby spectral features like the CH Λ-doublet mentioned previously.

During three flights in May 2016 the telescope was pointed towards NGC7027 (the total on-target integration time was 71 minutes). Weak emission in the HeH$^+$ $J = 1 – 0$ line has clearly been detected (Figure 1), as has emission from the nearby CH doublet. Notably, lines are well separated in frequency (Extended Data Figure 1). The velocity profile of the HeH$^+$ line matches nicely that of the excited CO $J =11 – 10$ transition, which was observed in parallel. The velocity-integrated line brightness temperature, $\int T_{mb}$ dv = 3.6 ±0.7 K km/s, corresponds to a line flux of 1.63 ×10$^{-13}$ erg s$^{-1}$ cm$^{-2}$. Because the 14.3" half-power beam response of upGREAT includes most of the NGC7027 ionized gas sphere, this result will be close to the total HeH$^+$ flux emitted in the $J = 1 – 0$ line. The flux is somewhat higher than the upper limit (1.26 ×10$^{-13}$ erg s$^{-1}$cm$^{-2}$) assigned to any residual HeH$^+$ contribution in the ISO observations, in the attempt[16] to separate the line from its blend with the CH doublet (see Extended Data Table 1 for the fluxes observed with upGREAT during this experiment).

We have modelled the HeH$^+$ abundance across NGC7027. We approximated the nebula as a constant-pressure, spherically-symmetric shell, and adjusted the pressure to obtain a Strömgren sphere of angular radius 4.6″, the geometric mean deduced from the 1.4 GHz radio continuum image[23]. We adopted a stellar luminosity of 1.0 ×10$^4$ $L_\odot$, a stellar effective temperature of 1.9 x 10$^5$ K[12], a source distance of 980 pc[11], and a He abundance of 0.12 relative to H. With the CLOUDY photoionization code[24] we calculated profiles of temperature and density (H, He$^+$ and e$^-$) as a function of position across the shell (Figure 2). The mean electron density within the ionized shell is 4.9 ×10$^4$ cm$^{-3}$ and ~2 ×10$^4$ cm$^{-3}$ in the H/He$^+$ overlap layer.

We then computed the equilibrium abundance of HeH$^+$, including the three reactions identified as

being important in the layers where HeH$^+$ is most abundant[7,13]

$$He^+ + H \rightarrow HeH^+ + h\nu \quad (R1)$$
$$HeH^+ + e^- \rightarrow He + H \quad (R2)$$
$$HeH^+ + H \rightarrow H_2^+ + He \quad (R3)$$

We confirm the conclusion previously reached[13] that in the planetary nebula environment, the reaction He + H$^+$ → HeH$^+$ + hν, which dominates HeH$^+$ formation in the early Universe, can be neglected, as well as the reaction H$_2^+$ + He → HeH$^+$ + H. Moreover, we confirm that the photodissociation of HeH$^+$ is slow compared to reactions (R2) and (R3) in the region where the HeH$^+$ emission arises, and can therefore also be neglected. For the reactions (R1) to (R3), we critically reviewed the most recent available estimates for the rate coefficients, as detailed in Extended Data Table 2. Values for the rate coefficient $k_1$, appearing in the literature for the radiative association reaction (R1) vary widely; here we adopt a value of 1.4 ×10$^{-16}$ cm$^3$s$^{-1}$, based on the most recently published cross-section[25]. An experimental study[27] of the dissociative recombination reaction (R2), involving measurements of the cross-section at energies up to 40 eV, derives values for the thermal rate coefficient that are plainly inconsistent with the cross-sections presented in that same study. A re-analysis of those measurements yields $k_2$ = 3.0 ×10$^{-10}$ cm$^3$s$^{-1}$ (at $T_{kin}$ = 10$^4$ K) – a value that is significantly smaller than that originally inferred from those measurements. To compute the emissivity of the HeH$^+$ $J = 1 – 0$ transition, we made use of recent estimates for the rate coefficients for electron impact excitation[28]. At the densities of relevance to NGC7027, collisional de-excitation can be neglected, and the $J = 1– 0$ emissivity is determined by the total rate of excitation from $J = 0$ to all states with $J > 0$, for which we obtain a value of 2.8 ×10$^{-7}$ ($T$/10$^4$ K)$^{-0.5}$ cm$^3$s$^{-1}$.

Given the rate coefficients discussed above, our model predicts an integrated main beam brightness temperature of 0.86 K km s$^{-1}$, a factor ~4 below the value we observe. The most uncertain of the rate coefficients is probably $k_1$; if we take the approach of adjusting its value to fit the observed line intensity, we obtain $k_1$ = 6.0×10$^{-16}$ cm$^3$s$^{-1}$. Figure 2 shows results obtained for this value. As expected, the production of HeH$^+$ peaks sharply in the He$^+$/H overlap layer, reaching a peak abundance relative to H nuclei of 4.0×10$^{-8}$. The column density, from the center to the edge of the nebula is 2.4 ×10$^{12}$ cm$^{-2}$, and the total $J$=1–0 model line flux emitted by NGC7027 is 2.1×10$^{-13}$ erg s$^{-1}$ cm$^{-2}$. For the infrared ro-vibrational lines, we compute fluxes that are consistent with previous non-detections. For the v=1–0 R(0) line, observed using a circular aperture of 8″ that did not fully encompass the source[14], the predicted flux of 1.3×10$^{-14}$ erg s$^{-1}$cm$^{-2}$ is a factor ~3 below the observed upper limit. For the v=1–0 P(2) line, the flux predicted in a 0.87×10.3″ slit is 5.9 × 10$^{-15}$ erg s$^{-1}$cm$^{-2}$ is comparable to the reported[15] upper limit of 5×10$^{-15}$ erg s$^{-1}$cm$^{-2}$.

A comparison of the observed $J = 1 – 0$ flux with predictions of our excitation model in the well-constrained physical environment of the NGC7027 nebula casts light on the relative importance of the molecule's various formation and destruction paths and, in particular, does constrain the radiative association (R1) and the dissociative recombination rate (R2). In view of the large discrepancies reported in some of the literature, and because a model based on the latest cross-sections underpredicts the observed line fluxes, our findings may stimulate further advanced studies of these important reactions (and of the corresponding radiative association of He and H$^+$ that dominates under conditions in the early universe). The validation of these uncertain values by our astronomical measurement is limited by the relative simplicity of our physical model for the source - as was the case in previous modelling efforts[12], we have approximated an elongated nebula as being spherically-symmetric. Future non-spherically-symmetric models have the potential to further refine our estimates, but are beyond the scope of the present study.

Although HeH$^+$ is of limited importance on Earth today, the chemistry of the Universe began with the helium hydride ion. The lack of definitive evidence of its very existence in interstellar space has been a dilemma for astronomy. The unambiguous detection reported here brings a decade-long search to a happy ending at last - a success that has become possible due to maturing terahertz technologies (incorporated in the upGREAT instrument) and the timely availability of the unique SOFIA observatory, which allows high-altitude flights above the absorbing layers of Earth' atmosphere.

**Acknowledgements.** upGREAT is a development by the MPI für Radioastronomie and the KOSMA/Universität zu Köln, in cooperation with the DLR Institut für Optische Sensorsysteme. The development of upGREAT is financed by the participating institutes, by the German Aerospace Center (DLR) under grants 50 OK 1102, 1103 and 1104, and within the Collaborative Research Centre 956, funded by the Deutsche Forschungsgemeinschaft (DFG). The work of D.N. was supported by grant No. 120364 from NASA's Astrophysical Data Analysis Program (ADAP).

SOFIA is jointly operated by the Universities Space Research Association, Inc. (USRA), under NASA contract NAS2-97001, and the Deutsches SOFIA Institut (DSI) under DLR contract 50 OK 0901 and 50 OK 1301 to the University of Stuttgart.

We thank the SOFIA operations and engineering teams, whose dedication and supportive response to our – in particular for this project - demanding requests has been essential for the accomplishments reported here. Thank you, Erick Young and Göran Sandell, for making this possible!

We are very grateful to Oldrich Novotny for re-computing the thermal rate coefficient for dissociative recombination of $HeH^+$, based on experimental merged-beam cross-section measurements in the literature[27]. We thank Jerome Loreau for providing published cross-section calculations[25] for the radiative association reaction (R1) in tabular form, and for clarifying that the published cross-sections apply specifically to collisions of H(1s) and $He^+$(1s) in the singlet state.

**Author Contributions.** R.G. has initiated and planned the observations. H.W. performed the calibration of the data. D.N. performed the astrochemical modelling. RG, HW, KM, DN wrote the text. Extending the reception bandwidth of the upGREAT receiver to frequencies beyond 2 THz has been a joint year-long effort by the GREAT team. All authors contributed to the interpretation of the data and commented on the final manuscript.

**Author Information.** Reprints and permissions information is available at www.nature.com/reprints. The authors declare no competing financial interests. Readers are welcome to comment on the online version of the paper. Correspondence and requests for materials should be addressed to R.G. (rguesten@mpifr-bonn.mpg.de).

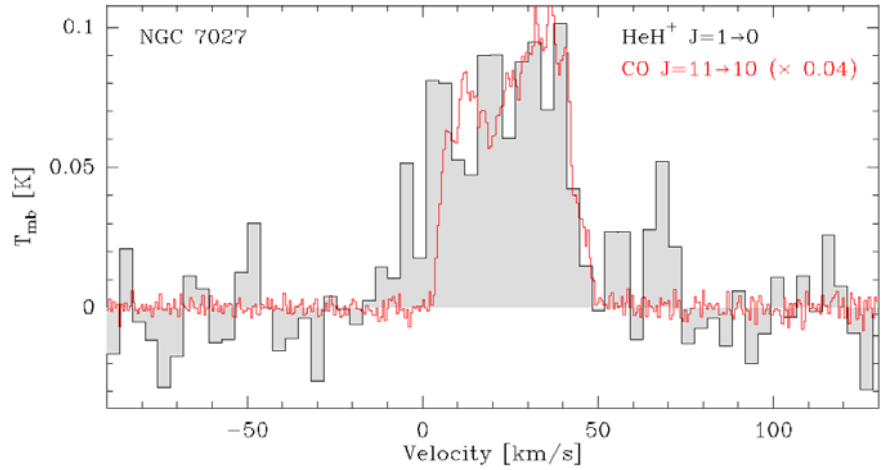

**Figure 1 | Spectrum of the HeH$^+$ ground-state rotational transition, observed with upGREAT on board of SOFIA towards NGC7027.** The "contaminating" emission from the nearby, but well separated CH Λ-doublet has been removed from the data (see Methods for details of the data processing). The spectrum has been re-binned to 3.6 km/s (24 MHz) resolution. For comparison, the CO $J = 11 - 10$ line is shown superimposed (at 0.58 km/s spectral resolution) - the transition was observed in parallel and probes the dense inner edge of the molecular envelope near the ionization front from where the HeH$^+$ emission is expected to originate.

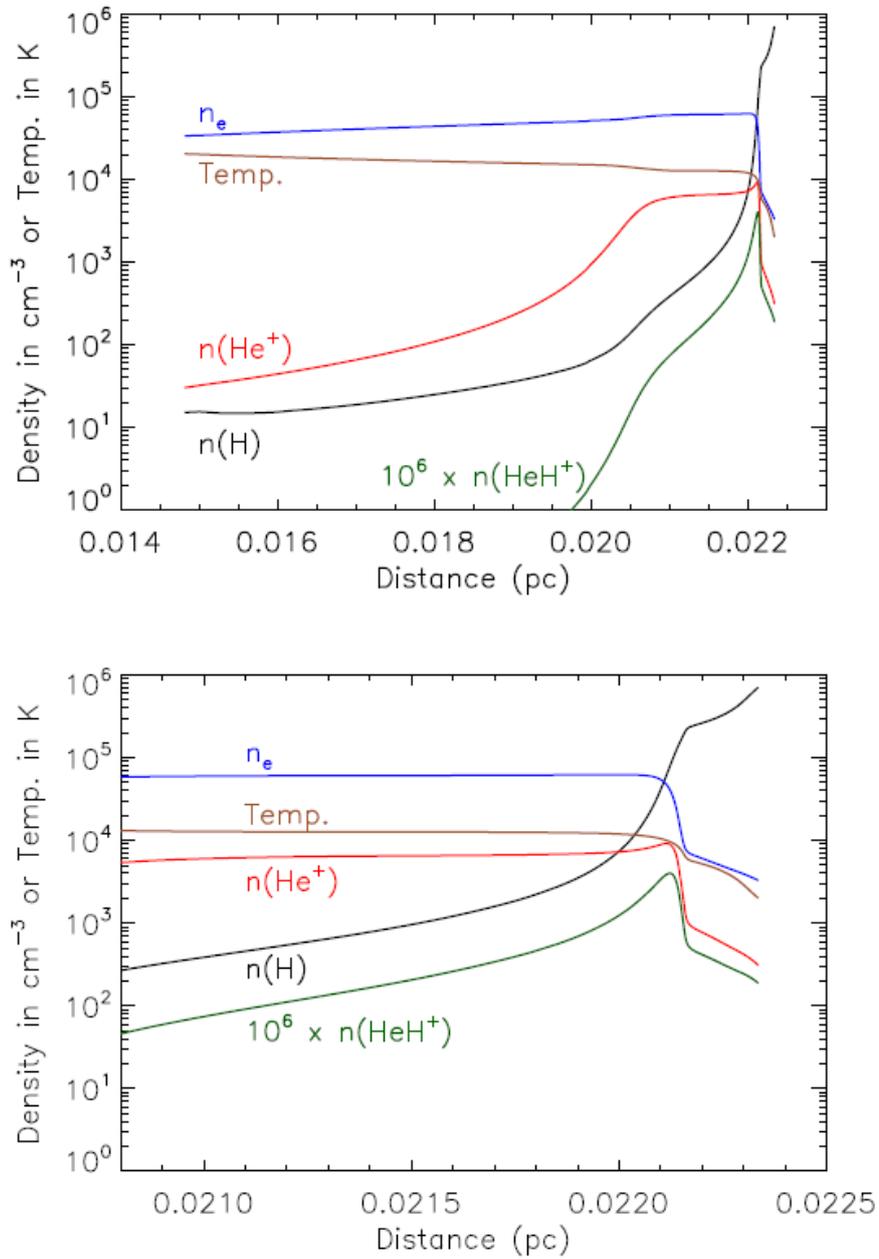

**Figure 2** | Temperature and density profiles for NGC 7027, as predicted by our astrochemical model for the source. The bottom panel presents an expanded view of the top panel. The model assumes a dissociative recombination rate (R2) of $k_2=3.0\times10^{-10}$ $(T/10^4$ K$)^{-0.47}$ cm$^3$s$^{-1}$ (see Extended Data Table 2) and a radiative association rate (R1) adjusted to fit the observed $J=1-0$ line intensity: $k_1=6.0\times10^{-16}$ cm$^3$s$^{-1}$.

# METHODS

**SOFIA/GREAT observations.** The planetary nebula NGC7027 (Right Ascension: $21^h07^m01.59^s$, Declination $+42°14'10.2''$, J2000) was observed with the GREAT[8] heterodyne spectrometer during three flights between 2016 May 14 and 25 out of Palmdale, USA. The 7-pixel array receiver upGREAT/LFA-H[9], tuned to the frequency of the HeH$^+$ $J = 1 – 0$ transition at 149.13 µm (2010.1839 GHz), was operated in parallel to the L1 channel recording the CO $J = 11 – 10$ transition for reference. Tuning the H-polarization of the array to the HeH$^+$ frequency has become possible with the installation of a new solid-state local oscillator reference source (developed by Virginia Diodes Inc.), just prior to this flight series. The NbN-based Hot Electron Bolometer mixers[29] of upGREAT operate at unmatched low noise temperatures of 850 K DSB, with a 3dB noise bandwidth of the intermediate frequency of 3.7 GHz. The signals were processed with our digital 4 GHz wideband monolithic FFT spectrometers[30], operating 16$k$ channels (with 244 kHz spectral resolution) for each of the front-end channels. Unless noted, all spectra presented in this contribution have been box-smoothed to 24.4 MHz spectral resolution as appropriate for the velocity width of the source.

The instrument was operated in double-beam chopped mode, with a beam throw of $40''$, at a chop rate of 2.5 Hz. Pointing was established by the telescope operators on nearby optical reference stars, to an accuracy of 1-2$''$. Prior to the flight series, the optical axis of the GREAT instrument had been aligned to these imagers by observations of Jupiter. The main beam coupling efficiencies for the L1 and upGREAT/LFA, also determined towards Jupiter, are 0.66 for both channels. The half-power beam widths (HPBW) of GREAT at 2010.18 and 1267.01 GHz are $14.3''$ and $21.1''$, respectively, diffraction-limited for the 2.5 m SOFIA telescope.

The observations were performed at flight altitudes between 40 and 43 kft; atmospheric conditions were typical for late spring flights out of Palmdale, CA, with a residual water vapour column of ~20 µm. This resulted in typical single-sideband system temperatures $T_{sys}$ of 1800 K.

**Observing Strategy**. As the upGREAT heterodyne spectrometer is sensitive to signals from both sidebands of the mixer, response from both mixing products $\nu_{LO} \pm \nu_{IF}$ is detected - where $\nu_{LO}$ is the frequency of the local oscillator reference signal, and $\nu_{IF}$ is the intermediate frequency (usable range 0.3 to 4 GHz). The HeH$^+$ observations were performed in the "upper sideband" (USB, $\nu_{LO} < \nu_{HeH+}$), because tuning the line into the lower sideband would pick-up excess noise from strong atmospheric emission in the image band. To exclude any chance coincidence with contamination from the image sideband, and thereby a faulty assignment of transitions, it is common practice in detection experiments to verify the assignments with observations of shifted IF reference signals. The first part of our observation was performed with an IF frequency offset of 1.4 GHz, the confirming observations during the 2$^{nd}$ and 3$^{rd}$ flight used an offset of 1.2 GHz (this shifted any feature from the image band by -60 km/s).

**Calibration and Analysis.** The calibration of the HeH$^+$ data is challenging. The line is affected by the proximity of a rather narrow ozone line at 2009.9 GHz which is optically thick at line center. Because the line originates in the signal band, the only way to minimize its impact is the seasonal variation of the Doppler correction. For May the line had moved out of the expected velocity range of NGC7027 - in autumn, however, it would be blending its core velocities, making this detection experiment impossible.

The spectrum was corrected for atmospheric losses following the usual calibration scheme[31] employing two load signals (at ambient and a cold temperature to determine the instrument gain) plus a blank sky signal which allows to fit an atmospheric model to the observed sky emission. The thus obtained transmission is subsequently used to correct the observed astronomical signal to the level it has outside the atmosphere. While the applied method proves adequate to correct for the broad absorption features owing to water vapour and collision-induced absorption by $N_2$ and $O_2$, it fails to provide a close fit to the prominent ozone line. We therefore deduced the absorption profile of this line directly from its observed emission, assuming that it mainly originates from a stratospheric layer with a constant source function.

**Data Analysis.** As mentioned previously, the high spectral resolution of upGREAT is essential to separate the HeH$^+$ $J = 1 - 0$ line from nearby transitions of the methylidyne radical, CH. HeH$^+$ is offset only by -619 MHz (+92 km/s) from the group of hyperfine transitions of the upper of the CH $^2\Pi_{1/2}$ J=3/2-1/2 Λ-doublet. With a Λ-splitting of ~4.05 GHz, the lower Λ-doublet will unavoidably blend from the image band into our spectrum. The frequency set-up for our observations was optimized prior to the actual flights, by simulating the blends of molecular transitions superposed on the atmospheric transmission. In Extended Data Figure 1 we display the calibrated "raw" spectra, as observed in the two frequency settings. The positions of the CH hyperfine transitions and of the HeH$^+$ $J = 1 - 0$ line are marked, corrected for the systemic source velocity of NGC7027 ($V_{lsr}$ = 26 km/s). The sky transmission is superimposed, displaying also the strong telluric (ozone) absorption at positive velocities.

**Data Availability.** The data presented in this contribution will be available through the SOFIA data archive at https://dcs.arc.nasa.gov/ and can be retrieved by searching for the project ID 83_0405.

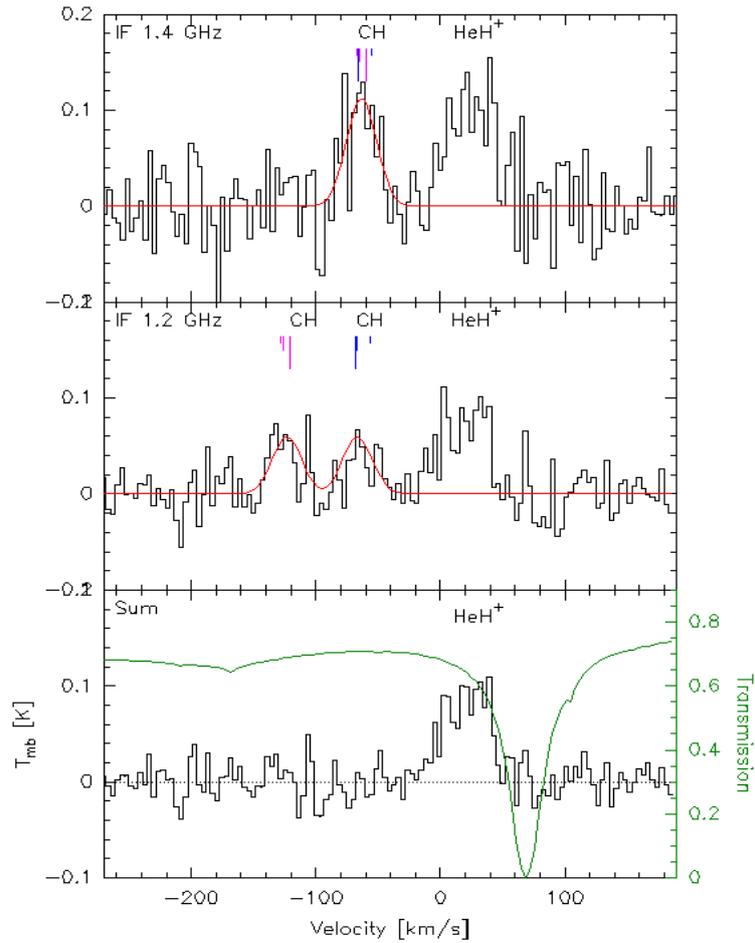

**Extended Data Figure 1 |** Calibrated and baseline corrected, but otherwise unprocessed spectra observed in two different frequency set-ups ($\nu_{IF}$ = 1.4 and 1.2 GHz, respectively, see text for details). The frequencies of the group of hyperfine transitions of the CH Λ-doublets are marked (blue: upper doublet from the signal band, purple: lower doublet blending from the image band). A pattern fit (optically thin, with intensities of the hyperfine pattern per Extended Data Table 1) is superposed with red lines. The bottom spectrum displays the co-added residuals of the two observations, after removal of the CH emission (shown is the residual after removal of the Gaussian fits). The atmospheric transmission is shown with a green line, for typical conditions encountered in May 2016 (precipitable water vapour column 20 μm, opacity of dry atmospheric constituents scaled by a factor of 1.4 with respect to the reference model, for a sightline at 35° elevation).

**Extended Data Table 1.** Molecular line parameters and line intensities as observed during this experiment.

| Transition $^2\Pi_J$ | hfs $F^p$ | $\nu_{hfs}$ [GHz] | $\lambda_{hfs}$ [μm] | relative line strength | line intensity [K km/s] |
|---|---|---|---|---|---|
| CH $^2\Pi_{3/2} \to {}^2\Pi_{1/2}$ | $1^- \to 1^+$ | 2006.7488646 | 149.392 | 0.2 | 1.5 (0.3) |
| | $1^- \to 0^+$ | 2006.7625778 | 149.391 | 0.4 | |
| | $2^- \to 1^+$ | 2006.7990641 | 149.308 | 1.0 | |
| | $1^+ \to 1^-$ | 2010.7385887 | 149.096 | 0.2 | 1.5 (0.3) |
| | $1^+ \to 0^-$ | 2010.8104600 | 149.090 | 0.4 | |
| | $2^+ \to 1^-$ | 2010.8119200 | 149.090 | 1.0 | |
| HeH$^+$ (J=1-0) | | 2010.1838730 | 149.137 | -- | 3.6 (0.7) |

Note: molecular line parameters have been taken from the Cologne Database for Molecular Spectroscopy[32], where the origin of the CH[33] and HeH$^+$ [34] frequencies is discussed. Velocity-integrated line intensities have been derived using the CLASS software package (http://www.iram.fr/IRAMFR/GILDAS). To the HeH$^+$ line we assign an absolute flux uncertainty of 20%, dominated by the challenging calibration at the edge of the strong, though narrow telluric absorption (see Extended Data Figure 2).

The CH doublets are fit using the relative strengths between the hyperfine transitions. For the decomposition of the two IF settings we assume equal intensity for the 2 groups of hyperfines, the different calibration in the two sidebands has been taken into account.

**Extended Data Table 2.** Reaction rates used in this contribution

| Reaction | [cm$^3$ s$^{-1}$] | Reference | Notes |
|---|---|---|---|
| He$^+$ + H $\to$ HeH$^+$ + h$\nu$ | $1.4 \times 10^{-16}$ | [25] | 1 |
| HeH$^+$ + e $\to$ He + H | $3.0 \times 10^{-10} T_4^{-0.47}$ | [26][27] | 2 |
| HeH$^+$ + H $\to$ He + H$_2^+$ | $1.2 \times 10^{-9} T_4^{-0.11}$ | [35] | |

(1) The rate coefficient was obtained by averaging the cross-sections[25] over a Maxwell-Boltzmann distribution, and applies for temperatures in the 5000 – 20000 K range. The published cross-section is for collisions where He$^+$(1s) and H(1s) are in the singlet state (total spin 0) [confirmed in communication with J.Loreau], but as only one quarter of collisions will have spin 0, the rate has been reduced by factor 4. Values in the literature[13,36,37] have varied widely.

(2) $T_4 = T_{kin}/10^4$ K. This value corrects an error in the computation[27] of the thermal rate coefficient from the cross-section measurements. Taking their primary data, the merged beams cross sections given in their Fig.3, as valid, the thermal rate coefficients were re-computed by applying the methods developed, e.g., for the calculation of the dissociative recombination cross-section of HCl$^+$ [26] - it is an order of magnitude below that originally computed[27].